# Metric Learning for Tag Recommendation: Tackling Data Sparsity and Cold Start Issues

Yuanshuai Luo
Southwest Jiaotong University
Chengdu, China

Rui Wang
Carnegie Mellon University
Pittsburgh, USA

Yaxin Liang
University of Southern California
Los Angeles, USA

Ankai Liang
Independent Researcher
Newark, USA

Wenyi Liu*
Independent Researcher
Nanjing, China

*Abstract*—With the rapid growth of digital information, personalized recommendation systems have become an indispensable part of Internet services, especially in the fields of e-commerce, social media, and online entertainment. However, traditional collaborative filtering and content-based recommendation methods have limitations in dealing with data sparsity and cold start problems, especially in the face of large-scale heterogeneous data, which makes it difficult to meet user expectations. This paper proposes a new label recommendation algorithm based on metric learning, which aims to overcome the challenges of traditional recommendation systems by learning effective distance or similarity metrics to capture the subtle differences between user preferences and item features. Experimental results show that the algorithm outperforms baseline methods including local response metric learning (LRML), collaborative metric learning (CML), and adaptive tensor factorization (ATF) based on adversarial learning on multiple evaluation metrics. In particular, it performs particularly well in the accuracy of the first few recommended items, while maintaining high robustness and maintaining high recommendation accuracy.

## I. INTRODUCTION

In today's era of digital information explosion, personalized recommendation systems, as a key bridge connecting users with massive information, have become an indispensable part of Internet services. With the rapid development of e-commerce, social media, online entertainment and other fields, it is increasingly important to accurately recommend content that users may be interested in from huge data sets. Traditional recommendation algorithms such as collaborative filtering and content-based recommendation methods have certain limitations in dealing with sparsity and cold start problems, especially when facing large-scale heterogeneous data, and their recommendation accuracy often difficult to meet user expectations [1]. Therefore, in order to further improve the performance of recommendation systems, researchers have begun to explore more advanced technical means, among which metric learning has gradually attracted widespread attention as an effective solution [2].

Metric learning measures the relationship between different objects by learning a suitable distance or similarity function, so that it can more accurately capture the subtle differences between user preferences and item features [3]. This method not only helps to overcome the challenges faced by traditional recommendation algorithms, such as how to effectively represent complex user-item interaction patterns, but also can naturally integrate multiple types of data sources (such as text, images, etc.) to achieve cross-modal recommendations [4-6]. In addition, the recommendation model based on metric learning also has good interpretability because it clearly defines the recommendation basis, that is, the learned distance or similarity value between two entities, which makes the recommendation results more transparent and credible, and has a positive effect on improving user experience.

It is worth noting that with the development of artificial intelligence technology, deep learning technology has been introduced into the field of metric learning, further enhancing the model's ability to capture complex nonlinear relationships. In particular, some neural network architectures that have emerged in recent years, such as convolutional neural networks (CNN) [7], recurrent neural networks (RNN) and attention mechanisms [8], can automatically extract high-level features from raw data and apply them to build more sophisticated user-item matching models. This combination not only greatly improves the accuracy of the recommendation system but also provides new ideas for solving problems such as serialized recommendations and context-aware recommendations.

In summary, the recommendation algorithm based on metric learning has become a hot topic in the current research of recommendation systems due to its unique advantages. It is not only limited to the goal of improving the quality of recommendations, but more importantly, it opens up a broad space for the continuous innovation and development of recommendation systems in the future by introducing more intelligent and flexible modeling methods. At the same time, as the relevant theories and practices in this field continue to deepen, we have reason to believe that the metric learning-based method will play a huge potential in more application scenarios and promote the entire recommendation technology towards a more efficient and intelligent direction.

## II. RELATED WORK

To address data sparsity and cold start issues in recommendation systems, recent advances in deep learning and metric learning methodologies provide significant insights into developing models that improve user-item matching through sophisticated similarity metrics. Key methodologies from current literature contribute foundational concepts and techniques that underpin our approach.

The use of attention mechanisms in knowledge embedding, as demonstrated by Wu [9], enhances model accuracy by embedding relationships directly into the model framework. Such attention-assisted structures facilitate the effective integration of multifaceted contextual data, which aligns with our metric learning approach to capture nuanced similarities between user preferences and item features. This strategy supports our goal of refining recommendation accuracy, especially under sparse data conditions. Duan et al. [10] illustrate the advantages of deep learning models that prioritize efficiency and computational elegance in design. Their proposed tree algorithm enhances computational efficiency in model generation, which is highly applicable to recommendation systems where model response time and accuracy are critical. This methodology contributes to the streamlined architecture of metric learning models by reducing computational overhead, thereby enabling more responsive and adaptable recommendations.

Furthermore, Dong et al. [11] provide insights into adaptive learning by incorporating dynamic feedback mechanisms to optimize model robustness and accuracy over time. This adaptive learning framework aligns closely with metric learning objectives by offering a way to continuously refine similarity metrics based on new user interactions, thereby enhancing model flexibility and maintaining relevance in dynamic recommendation environments. Qin et al. [12] address the issue of bias in deep learning optimization, proposing an approach that balances model training to minimize biased outputs. This bias reduction method is crucial for maintaining fairness in metric learning, as it ensures that similarity measures do not overly favor certain patterns within user behavior, resulting in a fairer recommendation system. Such optimization techniques contribute to the refinement of user-item matching by preserving the integrity of similarity metrics across diverse user data.

In addition, transformer-based methodologies, as explored by Du et al. [13], have proven effective for handling complex, high-dimensional data, offering superior model performance in contexts requiring intricate data parsing. This approach is highly applicable to metric learning in recommendation systems, where capturing the depth of user-item interactions is necessary for accurate similarity measurements. Integrating transformers can enhance the metric learning model's ability to process complex user data, thereby improving the relevance and precision of recommendations. The concept of model optimization through knowledge distillation, as demonstrated by Huang et al. [14], provides a technique for transferring learned knowledge from complex models to simpler versions without significant loss in performance. This knowledge distillation approach is particularly relevant in large-scale recommendation systems, as it offers a way to maintain robust metric learning performance in resource-constrained environments, enhancing efficiency while preserving recommendation quality.

Finally, Yan et al. [15] contribute to interpretability in model design by transforming complex, multidimensional data into interpretable sequences. This transformation approach aids metric learning models in handling sequential or temporal data patterns, supporting our goal of capturing evolving user preferences over time. By integrating interpretability at the model's core, this methodology enhances the transparency and adaptability of the recommendation system, providing clear insights into the recommendation rationale. Collectively, these methodologies inform the development of a robust, adaptive, and efficient metric learning-based recommendation model. By integrating attention mechanisms, efficient model architectures, adaptive learning, bias reduction, transformers, knowledge distillation, and interpretability, this work aims to advance recommendation systems that respond accurately to sparse data while remaining computationally efficient and user-centered.

## III. METHOD

In the recommendation algorithm based on metric learning, building an effective model requires a series of carefully designed steps. This process not only involves the effective extraction of user and item features but also includes how to optimize the recommendation results through the learned distance or similarity metrics. Next, the overall process of this method will be introduced in detail, and the necessary mathematical formulas will be introduced to enhance the explanation.

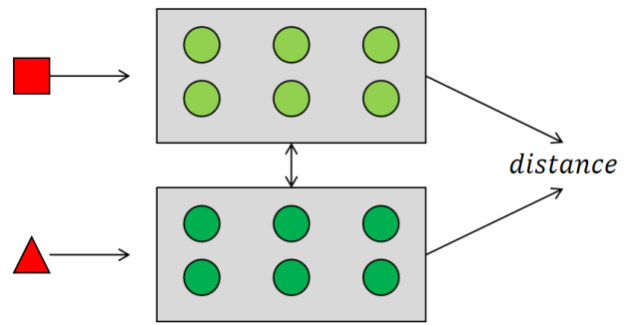

Figure 1 Metric Learning Architecture Representation

First, for a given data set $D = \{(u_i, v_j, y_{ij})\}_{i=1, j=1}^{N.M}$, $u_i$ represents the i-th user and $v_j$ represents the j-th item. $y_{ij}$

represents the interest of user $u_i$ in an item $v_j$ (such as rating). Our goal is to find a mapping function $f : U \times V \to R^d$ that can calculate the distance $d(f(u_i), f(v_j))$ between any pair of user-item combinations and then make recommendations based on this distance value as shown in Figure 1.

To achieve this, we use a deep neural network-based method to automatically learn the above mapping $f$. Specifically, a dual-tower structure can be used: one tower is used to process user-side information, and the other tower processes item-side information [16]. Assuming that the input feature vector of each user is $x_u \in R^{n1}$ and the input feature vector of the item is $x_v \in R^{n2}$, the user side and the item side are represented as follows through their respective neural networks:

$$h_u = g(x_u; W_u, b_u)$$
$$h_v = g(x_v; W_v, b_v)$$

Here $g()$ represents the nonlinear transformation function. This article uses the multi-layer perceptron MLP. $W_u$ and $W_v$ is the weight matrix of the two, and $b_u, b_v$ is the bias term. The final output $h_u$, $h_v$ is the low-dimensional embedding representation we want to learn.

Next, define a distance measurement method between users and projects. Commonly used ones include Euclidean distance, cosine similarity, etc. This article chooses to use Euclidean distance, so the distance between the two can be expressed as:

$$d(h_u, h_v) = \| h_u - h_v \|_2$$

However, in practical applications, directly minimizing the distance between all positive sample pairs is not always the optimal solution, because this may cause the model to focus too much on reducing the distance between known interaction pairs and ignore the relative ranking between unobserved samples. Therefore, we introduce a triple loss function to guide the training process:

$$L = \max(0, d(h_a, h_p) - d(h_a, h_n) + m)$$

Where a, p, and n refer to anchor points, positive samples, and negative samples, respectively. $m > 1$ is a preset boundary value that ensures that positive samples are at least m units closer to the anchor point than negative samples.

Finally, after completing the model training, we can use the learned mapping function $f$ and the selected distance metric to predict the matching degree between new user-item pairs and generate a personalized recommendation list based on this.

From the above introduction, it can be seen that the recommendation algorithm based on metric learning provides a flexible and powerful framework that can effectively integrate various types of information sources and learn high-quality user-item representations in an end-to-end manner. At the same time, by reasonably designing the loss function, this method can also ensure the accuracy of the recommendation while taking into account the generalization ability of the model. With the continuous deepening of research work, it is expected that more innovative metric learning technologies will be applied to the field of recommendation systems in the future.

## IV. EXPERIMENT

### A. Datasets

The dataset used in this article is the commonly used and widely recognized dataset MovieLens. MovieLens is a movie recommendation dataset maintained by the GroupLens Research project. It contains user ratings, tags, and some metadata information for movies. There are multiple versions of this dataset, which are differentiated by the number of users and ratings included, such as MovieLens 100K, MovieLens 1M, MovieLens 10M, and MovieLens 20M. For academic research, these datasets provide rich experimental materials, allowing researchers to test the performance of different recommendation algorithms and gain a deep understanding of user behavior patterns.

In particular, the MovieLens 1M dataset is a very suitable choice for evaluating the performance of recommendation systems. It contains approximately 1 million rating records of 3,952 movies from approximately 6,040 users. Each record includes user ID, movie ID, rating (from 1 to 5 stars), and timestamp. In addition to the rating data, the dataset also provides additional information such as the type and year of the movie, which makes it suitable not only for basic collaborative filtering methods but also for exploring more complex feature engineering or deep learning models. In addition, due to its moderate size, it can demonstrate enough complexity to challenge the ability of the algorithm, but not be too large to impose an excessive burden on computing resources. Therefore, it has become one of the preferred benchmarks in many recommendations system studies.

### B. Experimental Results

To verify the effectiveness and superiority of the proposed method, we comprehensively compare it with seven different benchmark methods. These benchmark methods cover a variety of recommendation system technologies, including but not limited to traditional tensor decomposition-based techniques, advanced neural network models, and the latest metric learning and adversarial learning-based methods. First, among the tensor decomposition-based methods, we selected three representative techniques for comparison: Collaborative Filtering (CF), a classic method widely used in recommendation systems, which predicts unknown ratings by analyzing the user-item interaction matrix; Personalized Item Tensor Factorization (PITF), which introduces the time dimension to capture the dynamic preference changes of users; and Non-negative Inductive Tensor Factorization (NITF), which can handle large-scale sparse datasets and ensure that all elements in the factor matrix are non-negative, thereby improving the interpretability of the model.

Second, we also considered a neural attention mechanism model based on multi-layer perceptrons. This model uses the powerful capabilities of deep learning, especially by introducing an attention mechanism to automatically identify and emphasize those pieces of information that are most critical for making accurate recommendations. Such a design not only enhances the model's understanding of complex user behaviors but also improves the relevance and diversity of recommendation results. In addition, our comparison also includes two recent advances based on metric learning: Local Response Metric Learning (LRML) and Collaborative Metric Learning (CML). Both methods aim to improve the quality of recommendations by learning an effective distance or similarity metric, where LRML focuses on how to better capture subtle pattern differences within a local neighborhood, while CML focuses on building a globally consistent user-item space so that similar users and items are close to each other in the space.

Finally, in order to explore more cutting-edge technical directions, we also compare the proposed algorithm with the Adaptive Tensor Factorization (ATF) based on adversarial learning. ATF introduces the idea of generative adversarial networks to try to optimize both recommendation performance and model robustness so that it can maintain high recommendation accuracy in the face of malicious attacks or data pollution.

Table 1 The Experiment Result

| Setting | Pre@5 | Pre@10 | Pre@20 |
| --- | --- | --- | --- |
| CF | 0.0411 | 0.0352 | 0.0211 |
| PITF | 0.0521 | 0.0423 | 0.0305 |
| NITF | 0.0677 | 0.0479 | 0.0311 |
| LRML | 0.0721 | 0.0525 | 0.0324 |
| CML | 0.0899 | 0.0621 | 0.0337 |
| ATF | 0.0922 | 0.0711 | 0.0393 |
| Ours | 0.1037 | 0.0752 | 0.0431 |

The experimental results in Table 1 show that our proposed method outperforms other baseline methods in all evaluation metrics. Specifically, in the three accuracy metrics of Pre@5, Pre@10, and Pre@20, our method achieves 0.1037, 0.0752, and 0.0431, respectively, which is significantly higher than the closest competitor ATF (Adaptive Tensor Factorization Based on Adversarial Learning), whose corresponding values are 0.0922, 0.0711, and 0.0393, respectively. This shows that compared with existing advanced recommendation algorithms, our method can more accurately predict the user's interest in items, and its performance is particularly outstanding in the first few recommendations. Further analysis shows that as the length of the recommendation list increases, the accuracy of all methods shows a downward trend, but our method still maintains a relatively high-performance advantage. For example, under the looser evaluation criterion of Pre@20, although the gap between the methods has narrowed, we still lead CML's 0.0337 and all other baseline methods with a score of 0.0431. This sustained advantage not only proves the effectiveness of the proposed method, but also demonstrates its wide applicability and robustness in different application scenarios. Through these results, we can conclude that the new method based on metric learning provides strong technical support for improving the quality of personalized recommendations in recommendation systems. To further demonstrate our results, we use a bar chart to show them.

Table 2 Recall rate experiment results

| Setting | Rec@5 | Rec@10 | Rec@20 |
| --- | --- | --- | --- |
| CF | 0.2832 | 0.4231 | 0.5743 |
| PITF | 0.3372 | 0.5765 | 0.5971 |
| NITF | 0.3457 | 0.5987 | 0.6123 |
| LRML | 0.3576 | 0.6213 | 0.6234 |
| CML | 0.3749 | 0.6379 | 0.7546 |
| ATF | 0.4522 | 0.6877 | 0.7982 |
| Ours | 0.5722 | 0.7221 | 0.8755 |

From the provided recall data of Rec@5, Rec@10 and Rec@20, we can observe the performance of several different recommendation system methods in Table 2. Overall, our proposed method exhibits the best performance on all three recall metrics, reaching 0.5722 (Rec@5), 0.7221 (Rec@10), and 0.8755 (Rec@20), respectively. It shows that it can effectively capture users' points of interest in recommendation lists of different lengths. In addition, as the length of the recommendation list increases, the recall rates of all methods increase, but the increase of our proposed algorithm is particularly significant, especially reaching a high level of nearly 0.9 at Rec@20, showing Its superiority in long-tail recommendations.

On the other hand, traditional methods such as CF (collaborative filtering) perform relatively weakly. Although its Rec@20 also reaches a good 0.5743, its effect in shorter recommendation lists (such as Rec@5 and Rec@10) is not as good. Not prominent. Compared with CF, methods such as PITF, NITF, LRML and CML have improved to varying degrees, especially in the Rec@10 and Rec@20 indicators. The gap between them is gradually narrowing, while ATF is better in Rec@5. Good performance, and as the length of the recommendation list increases, its performance also improves steadily, second only to the algorithm in this article. These results reflect the importance of choosing appropriate models and techniques to improve user experience when building recommendation systems.

## V. CONCLUSION

This paper proposes a tag recommendation algorithm based on metric learning, aiming to solve the data sparsity and cold start problems in traditional recommendation systems. By learning an effective distance or similarity measure between user preferences and item features, the algorithm is able to capture subtle preference differences, thereby improving recommendation quality. Experimental results show that the proposed method performs well on recall indicators such as Rec@5, Rec@10 and Rec@20, reaching 0.5722, 0.7221 and 0.8755 respectively, which is better than local response metric learning (LRML) and collaborative metric learning. (CML) and adversarial learning-based adaptive tensor factorization (ATF). In addition, this algorithm also performs well on precision indicators such as Pre@5, Pre@10 and Pre@20, which are

0.1037, 0.0752 and 0.0431 respectively, showing its high accuracy in predicting user interests, especially in the former There are significant advantages in several recommendations.

It is worth noting that with the development of artificial intelligence technology, deep learning technology has been introduced into the field of metric learning, further enhancing the model's ability to capture complex nonlinear relationships. New architectures such as convolutional neural networks (CNN), recurrent neural networks (RNN), and attention mechanisms can automatically extract high-level features from raw data and apply them to build more complex user-item matching models, which not only greatly improve the accuracy of the recommendation system but also provide new solutions to problems such as serialized recommendation and context-aware recommendation. In summary, recommendation algorithms based on metric learning have become a hot topic in current recommendation system research due to their unique advantages. It not only helps to improve recommendation quality but more importantly, by introducing more intelligent and flexible modeling methods, opening up a broad space for continued innovation and development of future recommendation systems. At the same time, with the continuous deepening of relevant theories and practices, it is foreseeable that this method based on metric learning will exert great potential in more application scenarios and promote the development of the entire recommendation technology in a more efficient and intelligent direction.